\documentclass[journal]{IEEEtran}
\usepackage{amsmath,graphicx}
\DeclareGraphicsExtensions{.pdf,.png,.jpg}

\ifCLASSINFOpdf
\else
\fi
\begin{document}
\title{Quantum Image Representation Through Two-Dimensional Quantum States and Normalized Amplitude}

\author{Madhur~Srivastava,~\IEEEmembership{Member,~IEEE,}
        Subhayan~Roy-Moulick,
        and~Prasanta~K.~Panigrahi% <-this % stops a space
\thanks{M. Srivastava is with the Department
of Biomedical Engineering, Cornell University, Ithaca,
NY, 14850 USA e-mail: ms2736@cornell.edu.}% <-this % stops a space
\thanks{S.Roy-Moulick and P.K. Panigrahi are with Department of Physical Sciences, Indian Institute of Science Education and Research Kolkata, Mohanpur, WB, 741246 India e-mail: subhayan@acm.org and pprasanta@iiserkol.ac.in.}% <-this % stops a space
\thanks{}}


\markboth{Submitted to IEEE Signal Processing Letters}%
{Srivastava \MakeLowercase{\textit{et al.}}: Bare Demo of IEEEtran.cls for Journals}



\maketitle

\begin{abstract}
We propose a novel method for image representation in quantum computers, which uses the two-dimensional (2-D) quantum states to locate each pixel in an image through row-location and column-location vectors for identifying each pixel location. The quantum state of an image is the linear superposition of the tensor product of the $m$-qubits row-location vector and the $n$-qubits column-location vector of each pixel. It enables the natural quantum representation of rectangular images that other methods lack. The amplitude/intensity of each pixel is incorporated into the coefficient values of the pixel's quantum state, without using any qubits. Due to the fact that linear superposition, tensor product and qubits form the fundamental basis of quantum computing, the proposed method presents the machine level representation of images on quantum computers. Unlike other methods, this method is a pure quantum representation without any classical components. 
\end{abstract}

\begin{IEEEkeywords}
Quantum Image Processing, Image Representation, Quantum Computing, Quantum Computer, Linear Superposition, Tensor Product.
\end{IEEEkeywords}

% \fi
\IEEEpeerreviewmaketitle

\section{Introduction}
\newcommand{\bra}[1]{\langle #1|}
\newcommand{\ket}[1]{|#1\rangle}
\newcommand{\braket}[2]{\langle #1|#2\rangle}

\IEEEPARstart{T}he idea of quantum computers, for the purpose of simulating quantum systems, was first proposed by Richard Feynman in 1982 [1]. In recent times, there has been significant progress in quantum computing. Quantum algorithms like Grover's database search [2] and Shor's integer factoring [3] have established the efficacy of quantum systems. The related fields of Quantum Information Processing, Quantum Information Theory, Quantum Cryptography and Quantum Communication have also been intensively explored [4-5]. Quantum Image Processing (QIP) is a relatively new addition to the quantum computation and information theory. QIP is currently at a nascent stage and its full scale development would enable various multimedia applications in quantum computers [6]. In comparison to classical image processing, QIP has exponential storage advantage i.e., where a square classical image requires $N \times N \times bitdepth$ bits, a quantum image needs only $(\lceil\log_2N\rceil + \lceil\log_2N\rceil + l)$ qubits [7-9]. The proposed method, two-dimensional quantum state and normalized amplitude (2-D QSNA), only requires $(\lceil\log_2N\rceil + \lceil\log_2N\rceil)$ qubits. 

In classical computers, image processing is a well-developed field, which makes use of a host of methods like Fourier, wavelet and other transforms for different applications. The recent developments of quantum Fourier transform [10-11] and quantum wavelet transform [12] have enabled the use of these transforms for various image processing applications on quantum computers. Images on the quantum computer need to be suitably represented, following the laws of quantum mechanics. For example, the image representation in classical computers is driven by the fact that all the information is represented in bits which can either be 0 or 1 at a given location and time. Quantum computers make use of qubits that obey the linear superposition principle. More precisely, the qubits are orthonormal vectors $\ket{0}=
\begin{pmatrix}
1 \\
0
\end{pmatrix}$ and $\ket{1}=
\begin{pmatrix}
0 \\
1
\end{pmatrix}$, or $\bra{0}=
\begin{pmatrix}
1 & 0
\end{pmatrix}$ and $\bra{1}=
\begin{pmatrix}
0 & 1
\end{pmatrix}$. A generic state can be represented with a superposition: $\ket{\psi} = \alpha_1 \ket{0} + \alpha_2 \ket{1}$ or $\bra{\psi} = \alpha_1 \bra{0} + \alpha_2 \bra{1}$, respectively, where $\alpha_1^2 + \alpha_2^2 = 1$, with $\alpha_1$ and $\alpha_2$ being real values (for images). To generate higher dimension vectors, the recursive tensor products (also called direct products; $\otimes$) of qubits can be taken. For instance, a 2-qubit vector is obtained  from the single qubit constituent vectors as follows, \begin{equation} \nonumber 
\ket{0} \otimes \ket{0} = \ket{00} = \begin{pmatrix}
1 \\
0 
\end{pmatrix} \otimes \begin{pmatrix}
1 \\
0 
\end{pmatrix} = \begin{pmatrix}
1 \\
0 \\
0 \\
0
\end{pmatrix}
\end{equation}

\noindent Similarly, $\ket{01} = \begin{pmatrix}
0 \\
1 \\
0 \\
0
\end{pmatrix}$, $\ket{10} = \begin{pmatrix}
0 \\
0 \\
1 \\
0
\end{pmatrix}$, $\ket{11} = \begin{pmatrix}
0 \\
0 \\
0 \\
1
\end{pmatrix}$, $\bra{00} = \begin{pmatrix}
1 & 0 & 0 & 0
\end{pmatrix}$, $\bra{01} = \begin{pmatrix}
0 & 1 & 0 & 0
\end{pmatrix}$, $\bra{10} = \begin{pmatrix}
0 & 0 & 1 & 0
\end{pmatrix}$, and $\bra{11} = \begin{pmatrix}
0 & 0 & 0 & 1
\end{pmatrix}$.

Venegas-Andraca and Bose [13] introduced image representation on the quantum computers by proposing the `qubit lattice' method, in which each pixel was represented in its quantum state and then a quantum matrix was created with them. The 'qubit lattice' representation was incorporated by Yuan et al. [14] in their simple quantum representation (SQR) method for infrared images. The SQR method replaced the color information with the radiation values as the coefficient values. Inspired by `qubit lattice', Li et al. [15] proposed a quantum representation of images which explicitly included and encoded the pixel position along with the color information. Subsequently, Li et al. [16-17] extended their previous works to multidimensional color images using quantum superposition. However, these methods [15-17] are constrained by qubit angle that has upper bound for the number of values it can possess. The qubit angle encodes the color information and is highly dependent on the image dimensions and the bit depth of color. In another work, Venegas-Andraca and Ball [18] proposed an  `entangled image' method for representing shapes in binary images through quantum entanglement. They only concentrated on binary images, whereas real life images possess multiple intensity levels. Both `qubit lattice' and `entangled image' are the quantum analog of classical images, and do not utilize the superposition property of quantum computation to represent all the pixels together. Latorre [19] proposed the `real let' approach that used quadtree to locate each pixel using 4-D qubit sequence. In order to be efficient, `real ket' requires image pixel values to be random, which is rare as images are highly correlated. Le et al. [7, 20] provided a flexible representation of quantum images (FRQI) for multiple intensity levels in a 2-D pixel representation, enabling various image processing operations and applications. Sun et al. [8,21] expanded  FRQI into three color channel RGB$\alpha$ image. Through novel enhanced quantum representation (NEQR), Zhang et al. [9] (also independently proposed by Caraiman and Manta [22]) provided an alternate approach to FRQI by storing the intensity information into qubits, along with the pixel information at the cost of increasing the number of qubits. Moreover, this method can only represent images with unsigned integer values. In a separate work, Zhang et al. [23] also presented a quantum image representation method, named quantum log-polar image (QUALPI), for the unsigned integer images acquired in the log-polar coordinate system. Among all the above methods, FRQI and NEQR are most comprehensive and have been used to develop many image processing applications and operations like image segmentation [15-16,18,24-25], geometrical transformations [26], color transformations [27], image similarity measures [28], encryption [29], watermarking [30-33], compression [7,9,17,20,22], image complement computation [9,22], image Binarization with a fixed threshold [9] and variable threshold [22], filtering [34], and histogram equalization [22] and computation [24,35]. A detailed literature survey can be found in [6,36].

%\begin{equation}
%\ket{Y} = \frac{1}{2^m} \displaystyle\sum_{I=0}^{2^m-1} \displaystyle\sum_{J=0}^{2^m-1} (\cos \theta_{IJ} \ket{0} + \sin \theta_{IJ} \ket{1}) \ket{IJ}
%\end{equation}
%\noindent where $\ket{Y}$ is the $2^m \times 2^m$ quantum image, and $\theta_{IJ}$ and $\ket{IJ}$ are the phase value of the intensity and the qubit state at the pixel $(I,J)$, respectively. 

%\begin{equation}
%\ket{Y} = \frac{1}{2^m} \displaystyle\sum_{I=0}^{2^m-1} \displaystyle\sum_{J=0}^{2^m-1} %\ket{f(I,J)} \ket{IJ}
%\end{equation}
%\noindent where $\ket{f(I,J)}$ is the qubit state of the unsigned integer gray-scale value at the pixel $(I,J)$.

However, there are two major drawbacks with the current state-of-art methods i.e., FRQI and NEQR. First, they are designed only for square image dimensions ($2^m \times 2^m$), although most of the images are rectangular ($2^m \times 2^n$). At present, $2m+l$ ($m$ qubits for each dimension) qubits are required to represent any quantum image. Second, they require additional $l$ qubits to represent pixel's amplitude or intensity value, resulting in the increase of storage space as well as reducing the compactness of representation. Also, with more qubits, the quantum states are prone to decoherence. 

To overcome these two drawbacks, here we propose a new quantum image representation (2-D QSNA) that allows the images to possess rectangular dimensions and represents the amplitude/intensity of a pixel without using any additional qubits. Hence, $m+n$ qubits without additional $l$ qubits will be required to represent quantum image. Considering the fact that the gray scale images or the different color channels of images are a 2-D matrix $Y$, with elements $Y_{p,q}$ having row and column indices $p$ and $q$, respectively. The method uses a dual representation of the row-location vector and column-location vector, separately, to identify a pixel in the image. Each location vector can be defined as the quantum state at $p^{th}$ row and $q^{th}$ column using multiple qubits. The tensor product of the row-location and column-location vector results in a 2-D quantum state of a pixel in the Liouvillian space. The pixel's 2-D quantum state is represented by multiple qubits, consisting of both row and column information. This dual representation permits rectangular dimensions of images as $I$ and $J$ can possess separate vector lengths. On the other hand, the pixel amplitude/intensity values are normalized and assigned as the scalar amplitude of the pixel's 2-D quantum state. The normalization of amplitudes is derived from the constraint that the scalar amplitudes of the quantum states should be an unit vector. Using scalar amplitude values to represent pixel amplitude does not require any additional qubits. In FRQI and NEQR methods (also QUALPI in log-polar domain), the scalar amplitudes are not used for storing any image information, making them redundant.

\section{Proposed Quantum Image Representation}

\subsection{Two-Dimensional Quantum State of Pixel Location}
\label{sec:1}
\textbf{The first new feature} of this paper is the dual representation of a 2-D image by the row-location and column-location vectors. It is achieved by generating a $M$-length row-location vector and a $N$-length column-location vector with $m$-qubits and $n$-qubits, respectively, where $m = \log_2 M$ and $n = \log_2 N$.  For the row-location vector, $m$-qubits representation can be generated with recursive tensor products ($\otimes$) of the single qubit constituent vectors, as mentioned earlier. Mathematically, it is defined as,
\begin{equation}
\ket{I}_p = \ket{i}^{\otimes m}
\end{equation}
\noindent where $i \in \{0,1\}$, $p \in [1,M]$, and $\ket{I}_p$ is the row-location vector or the quantum state of $m$-qubit at $p^{th}$ row.

Likewise, the column-location vector is generated by,
\begin{equation}
\bra{J}_q = \bra{j}^{\otimes n}
\end{equation}
\noindent where $j \in \{0,1\}$, $q \in [1,N]$, and $\bra{J}_q$ is the column-location vector or the quantum state of $n$-qubit at $q^{th}$ column.

To represent a pixel location in its 2-D matrix and to identify 2-D location of a pixel, the tensor product of the row-location vector (equation 1) and the column-location vector (equation 2) is carried out as follows,
\begin{equation}
L_{p,q} = \ket{I}_p \otimes \bra{J}_q
\end{equation}
\noindent where $L_{p,q}$ is the 2-D quantum state of a pixel at $p^{th}$ row and $q^{th}$ column using $m$-qubits and $n$-qubits, respectively.

To illustrate with an example, a pixel location in terms of 2-D quantum states using the row-location and column-location vector for any $4\times 4$ and $2\times 4$ image matrix is given by the equations 4 and 5, respectively.  
\begin{equation} \small
L_{p,q}^{4\times 4} \in
\begin{pmatrix}
\ket{00}\otimes \bra{00} & \ket{00}\otimes \bra{01} & \ket{00}\otimes \bra{10} & \ket{00}\otimes \bra{11}\\
\ket{01}\otimes \bra{00} & \ket{01}\otimes \bra{01} & \ket{01}\otimes \bra{10} & \ket{01}\otimes \bra{11}\\
\ket{10}\otimes \bra{00} & \ket{10}\otimes \bra{01} & \ket{10}\otimes \bra{10} & \ket{10}\otimes \bra{11}\\
\ket{11}\otimes \bra{00} & \ket{11}\otimes \bra{01} & \ket{11}\otimes \bra{10} & \ket{11}\otimes \bra{11}
\end{pmatrix}
\end{equation}
\begin{equation} 
L_{p,q}^{2\times 4} \in
\begin{pmatrix}
\ket{0}\otimes \bra{00} & \ket{0}\otimes \bra{01} & \ket{0}\otimes \bra{10} & \ket{0}\otimes \bra{11}\\
\ket{1}\otimes \bra{00} & \ket{1}\otimes \bra{01} & \ket{1}\otimes \bra{10} & \ket{1}\otimes \bra{11}
\end{pmatrix}
\end{equation}

\subsection{Normalization of Image Amplitudes}
\label{sec:2}
\textbf{The second new feature} of this paper is the incorporation of pixel amplitude/intensity values into the scalar amplitude of its respective 2-D quantum state, requiring no additional qubits. In the quantum computation theory [4-5], the scalar amplitudes ($\alpha$) of the quantum states, like the quantum states and the superposition of quantum states, are also constrained to be an unit vector, i.e.
\begin{equation}
\displaystyle\sum_{p=1}^M \displaystyle\sum_{q=1}^N \alpha_{p,q}^2 = 1
\end{equation}
\noindent where $\alpha_{p,q}$ is the scalar amplitude of the pixel quantum state at $p^{th}$ row and $q^{th}$ column.

Let $A_{p,q}$ be the amplitude/intensity of the pixel at $p^{th}$ row and $q^{th}$ column. To incorporate $A_{p,q}$ into $\alpha_{p,q}$ such that equation 6 is satisfied, $\alpha_{p,q}$ can be written as,
\begin{equation}
\alpha_{p,q} = \sqrt{\frac{A_{p,q}}{\displaystyle\sum_{p=1}^M \displaystyle\sum_{q=1}^N A_{p,q}}}
\end{equation}
As can be seen, $A_{p,q}$ is normalized with respect to total amplitude value $A_T^{M\times N}=\displaystyle\sum_{p=1}^M \displaystyle\sum_{q=1}^N A_{p,q}$, in order to be used as $\alpha_{p,q}$. It must be noted that $A_{p,q}$ can possess any value, whether unsigned integer or real values for gray scale images and the individual channels of color images. 

For example, the scalar amplitudes for 2-D quantum state for each pixel in images with $4 \times 4$ and $2 \times 4$ dimensions are shown in equations 8 and 9, respectively, 
\begin{equation}
\alpha^{4\times 4}_{p,q}\in \frac{1}{\sqrt{A^{4\times 4}_{T}}}
\begin{pmatrix}
\sqrt{A_{1,1}} &\sqrt{A_{1,2}} & \sqrt{A_{1,3}} & \sqrt{A_{1,4}} \\
\sqrt{A_{2,1}} &\sqrt{A_{2,2}} & \sqrt{A_{2,3}} & \sqrt{A_{2,4}} \\
\sqrt{A_{3,1}} &\sqrt{A_{3,2}} & \sqrt{A_{3,3}} & \sqrt{A_{3,4}} \\
\sqrt{A_{4,1}} &\sqrt{A_{4,2}} & \sqrt{A_{4,3}} & \sqrt{A_{4,4}} 
\end{pmatrix}
\end{equation}
\begin{equation}
\alpha^{2\times 4}_{p,q}\in \frac{1}{\sqrt{A^{2\times 4}_{T}}}
\begin{pmatrix}
\sqrt{A_{1,1}} &\sqrt{A_{1,2}} & \sqrt{A_{1,3}} & \sqrt{A_{1,4}} \\
\sqrt{A_{2,1}} &\sqrt{A_{2,2}} & \sqrt{A_{2,3}} & \sqrt{A_{2,4}}
\end{pmatrix}
\end{equation}

\subsection{Image Representation}
We now incorporate these structures for complete representation as an image. Making the use the 2-D quantum state representation of each pixel and its scalar amplitude, the quantum images can be represented as the superposition of all the pixel's quantum states along with their scalar amplitudes. The proposed quantum image representation, 2-D QSNA, is defined as,
\begin{equation}
Y = \displaystyle\sum_{p=1}^M \displaystyle\sum_{q=1}^N \alpha_{p,q} L_{p,q}
\end{equation}
\noindent where $Y$ is the 2-D quantum image. After substituting $L_{p,q}$ from equation 3 and $\alpha_{p,q}$ from equation 7, the above can be rewritten as,
\begin{equation}
Y = \frac{1}{\sqrt{A^{M\times N}_{T}}} \displaystyle\sum_{p=1}^M \displaystyle\sum_{q=1}^N \sqrt{A_{p,q}} (\ket{I}_p \otimes \bra{J}_q)
\end{equation}

\begin{figure}[t!]
\centering
 \includegraphics[width=0.47\textwidth]{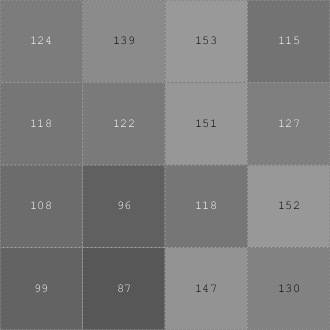}
\caption{A $4\times 4$ image with its intensity values}
\end{figure}

\section{Image Preparation, Retrieval, Example, and Comparison}

Unlike the current methods, the quantum images through 2-D QSNA would be prepared in the pure state i.e., through the experiemental technique and mechanism acquiring the image. These quantum images can be restored and retrieved using quantum tomographic procedures [37] or weak measurements [38-41] without collapsing the quantum state [42-43]. The other methods prepare the images from the pure state. The exact preparation and retrieval of 2-D QSNA requires rigorous theoretical and experimental quantum mechanics, which is a separate research work.

\newcounter{tempequationcounter}
\begin{figure*}[!t] 
\scriptsize
\setcounter{tempequationcounter}{\value{equation}}
\begin{equation} \nonumber 
Y^{4 \times 4} = \frac{1}{\sqrt{A^{4\times 4}_{T}}} \left[ \sqrt{A_{1,1}} (\ket{00} \otimes \bra{00}) + \sqrt{A_{1,2}} (\ket{00} \otimes \bra{01}) + \sqrt{A_{1,3}} (\ket{00} \otimes \bra{10}) + \sqrt{A_{1,4}} (\ket{00} \otimes \bra{11}) + \sqrt{A_{2,1}} (\ket{01} \otimes \bra{00})  \right]
\end{equation}
\begin{equation} \nonumber 
 + \frac{1}{\sqrt{A^{4\times 4}_{T}}} \left[ \sqrt{A_{2,2}} (\ket{01} \otimes \bra{01}) + \sqrt{A_{2,3}} (\ket{01} \otimes \bra{10}) + \sqrt{A_{2,4}} (\ket{01} \otimes \bra{11}) + \sqrt{A_{3,1}} (\ket{10} \otimes \bra{00}) + \sqrt{A_{3,2}} (\ket{10} \otimes \bra{01}) + \sqrt{A_{3,3}} (\ket{10} \otimes \bra{10})  \right]
\end{equation} 
\begin{equation}
 + \frac{1}{\sqrt{A^{4\times 4}_{T}}} \left[\sqrt{A_{3,4}} (\ket{10} \otimes \bra{11}) + \sqrt{A_{4,1}} (\ket{11} \otimes \bra{00}) + \sqrt{A_{4,2}} (\ket{11} \otimes \bra{01}) + \sqrt{A_{4,3}} (\ket{11} \otimes \bra{10}) + \sqrt{A_{4,4}} (\ket{11} \otimes \bra{11}) \right]
\end{equation}
\setcounter{equation}{\value{tempequationcounter}}
\hrulefill
\vspace*{4pt}
\end{figure*}
\addtocounter{equation}{1}

\begin{figure*}[!t] 
\scriptsize
\setcounter{tempequationcounter}{\value{equation}}
\begin{equation} \nonumber
Y^{2 \times 4} = \frac{1}{\sqrt{A^{2\times 4}_{T}}} \left[ \sqrt{A_{1,1}} (\ket{0} \otimes \bra{00}) + \sqrt{A_{1,2}} (\ket{0} \otimes \bra{01}) + \sqrt{A_{1,3}} (\ket{0} \otimes \bra{10}) + \sqrt{A_{1,4}} (\ket{0} \otimes \bra{11}) + \sqrt{A_{2,1}} (\ket{1} \otimes \bra{00})   \right]
\end{equation}
\begin{equation}
 + \frac{1}{\sqrt{A^{2\times 4}_{T}}} \left[ \sqrt{A_{2,2}} (\ket{1} \otimes \bra{01}) + \sqrt{A_{2,3}} (\ket{1} \otimes \bra{10}) + \sqrt{A_{2,4}} (\ket{1} \otimes \bra{11}) \right]
\end{equation} 
\setcounter{equation}{\value{tempequationcounter}}
\hrulefill
\vspace*{4pt}
\end{figure*}
\addtocounter{equation}{1}

\begin{figure*}[!t] 
\scriptsize
\setcounter{tempequationcounter}{\value{equation}}
\begin{equation} \nonumber 
Y^{4 \times 4} = \frac{1}{\sqrt{1986}} \left[\sqrt{124} (\ket{00} \otimes \bra{00}) + \sqrt{139} (\ket{00} \otimes \bra{01}) + \sqrt{153} (\ket{00} \otimes \bra{10}) + \sqrt{115} (\ket{00} \otimes \bra{11}) + \sqrt{118} (\ket{01} \otimes \bra{00}) + \sqrt{122} (\ket{01} \otimes \bra{01}) \right]
\end{equation}
\begin{equation} \nonumber
+ \frac{1}{\sqrt{1986}} \left[\sqrt{151} (\ket{01} \otimes \bra{10}) + \sqrt{127} (\ket{01} \otimes \bra{11}) + \sqrt{108} (\ket{10} \otimes \bra{00}) + \sqrt{96} (\ket{10} \otimes \bra{01}) + \sqrt{118} (\ket{10} \otimes \bra{10}) + \sqrt{152} (\ket{10} \otimes \bra{11})\right]
\end{equation}
\begin{equation}
+ \frac{1}{\sqrt{1986}} \left[ \sqrt{99} (\ket{11} \otimes \bra{00}) + \sqrt{87} (\ket{11} \otimes \bra{01}) + \sqrt{147} (\ket{11} \otimes \bra{10}) + \sqrt{130} (\ket{11} \otimes \bra{11})\right] 
\end{equation}
\setcounter{equation}{\value{tempequationcounter}}
\hrulefill
\vspace*{4pt}
\end{figure*}
\addtocounter{equation}{1}

\begin{figure*}[!t]
\scriptsize
\setcounter{tempequationcounter}{\value{equation}}
\begin{equation} \nonumber
Y^{2 \times 4} = \frac{1}{\sqrt{1049}} \left[\sqrt{124} (\ket{0} \otimes \bra{00}) + \sqrt{139} (\ket{0} \otimes \bra{01}) + \sqrt{153} (\ket{0} \otimes \bra{10}) + \sqrt{115} (\ket{0} \otimes \bra{11}) + \sqrt{118} (\ket{1} \otimes \bra{00}) \right]
\end{equation}
\begin{equation}
+ \frac{1}{\sqrt{1049}} \left[ \sqrt{122} (\ket{1} \otimes \bra{01}) + \sqrt{151} (\ket{1} \otimes \bra{10}) + \sqrt{127} (\ket{1} \otimes \bra{11}) \right]
\end{equation}
\setcounter{equation}{\value{tempequationcounter}}
\hrulefill
\vspace*{4pt}
\vspace{-4mm}
\end{figure*}
\addtocounter{equation}{1}

Equations 12 and 13 show the general form of the proposed representation for any $4 \times 4$ and $2 \times 4$ image, respectively. An example is also presented using a $4\times 4$ unsigned integer gray scale image (see Figure 1) in equation 14; however, it is not limited to just unsigned integer intensity values and can possess any real values. To show an example for a rectangular image, upper $2\times 4$ dimension of Figure 1 is considered. The quantum image of it is presented in equation 15. 

The key information to be observed is the number of qubits required to individually represent the row-location and column-location vectors for a $4\times 4$ image (equations 12 and 14) and a $2\times 4$ image (equations 13 and 15). A $4\times 4$ image requires $2$-qubits ($\ket{00}$,$\ket{01}$,$\ket{10}$, and $\ket{11}$), each for the row-location and column-location vector. On the other hand, an image with $2\times 4$ dimension requires $1$-qubit ($\ket{0}$ and $\ket{1}$) for the 
row-location vector and $2$-qubits ($\ket{00}$,$\ket{01}$,$\ket{10}$, and $\ket{11}$) for the column-location vector. The independent quantum states of the row-location and column-location vectors allow separate allocation of the number of qubits to each of them. Hence, 2-D QSNA is able to achieve the quantum representation for both square and rectangular images.

\begin{table}[!t]
%\renewcommand{\arraystretch}{1.3}
% if using array.sty, it might be a good idea to tweak the value of
% \extrarowheight as needed to properly center the text within the cells
\caption{Comparison of FRQI, NEQR, and the 2-D QSNA method.}
\label{table_example}
\centering
% Some packages, such as MDW tools, offer better commands for making tables
% than the plain LaTeX2e tabular which is used here.
\begin{tabular}{|c||c|c|c|c|}
\hline
Method & FRQI & NEQR & 2-D QSNA \\
\hline
Dimension & $2^m \times 2^m$ & $2^m \times 2^m$ & $2^m \times 2^n$ \\
Qubits Required & $2m+1$ & $2m + l$ & $m + n$ \\
Preparation Complexity & $O(2^{4m})$ & $O(2^{m})$ & Pure State \\
Image Format & Real & Unsigned Integer & Real \\
Intensity Storage & 1 qubit & $l$ qubits & None \\
\hline
\end{tabular}
%\vspace{-7mm}
\end{table}

Table I compares 2-D QSNA with FRQI and NEQR in terms of the types and dimensions of quantum images that can be represented as well as the number of qubits (i.e., storage) and complexity required to achieve it. Other methods are not compared as FRQI and NEQR are most widely used and have already shown to be comprehensive and better in previous literatures [6,25,36]. It can be seen through Table I that 2-D QSNA enables the quantum representation for a rectangular image and provides a general form for images with any dimensions, while FRQI and NEQR are only devised for square images. In addition, there is no preparation complexity for 2-D QSNA as it will be acquired in the pure state. The other two comprehensive methods are independent of the physical system that will prepare the quantum state, resulting in additional burden of preparation complexity. On the other hand, 2-D QSNA allows the physical system to design it own preparation. This freedom would be necessary and important for the specialized quantum imaging equipments like quantum camera and others.  Further, as shown in equation 11, the incorporation of pixel amplitude/intensity into the scalar amplitude of pixel's 2-D quantum state does not require any qubits, and hence, is independent of bit depth. The storage space for the proposed method would only depend on the image dimensions. The other methods used additional qubit/s to represent pixel amplitude/intensity. Hence, 2-D QSNA is less prone to decoherence as decoherence in quantum system is highly dependent upon the number of qubits. Lastly, due to the normalized amplitude scalar values, 2-D QSNA is valid for any color channel of any color space, allowing easy extension to multi-channel color image. 

\section{Conclusion}

The proposed 2-D QSNA method requires the minimum number of qubits and can represent both rectangular and square images in a pure quantum form, while other methods lack one or more features. This representation can throw light on nature of entanglement and correlation of the underlying images. It will also enable further manipulation like watermarking and cryptography. The dual representation of row and column makes extracting and projecting specific features of an image simpler. The future work will focus on developing quantum image storing and retrieving techniques, image compression techniques using the row-location and column-location quantum states, and operations for various image applications.

\bibliographystyle{IEEEtran}

\end{document}